# Gas-phase synthesis of carbon nanotube-graphene heterostructures

Saeed Ahmad[1], Hua Jiang[1], Kimmo Mustonen[2], Qiang Zhang[1*], Aqeel Hussain[1], Abu Taher Khan[1], Nan Wei[1], Mohammad Tavakkoli[1], Yongping Liao[1], Er-Xiong Ding[1], Jani Kotakoski[2], and Esko I. Kauppinen[1*]

[1]Department of Applied Physics, Aalto University School of Science, P.O. Box 15100, FI-00076 Aalto, Finland.

[2]Faculty of Physics, University of Vienna, Boltzmanngasse 5, 1090 Vienna, Austria.

**Abstract**

Graphene and carbon nanotubes (CNTs) share the same atomic structure of hexagonal carbon lattice. Yet, their synthesis differs in many aspects, including the shape and size of the catalyst. Here, we demonstrate a floating-catalyst chemical vapor deposition (FCCVD) technique for substrate-free, single-step growth of CNT-graphene heterostructures (CNT-G-H) using ethylene as a carbon source. The formation of CNT-G-H is directly evidenced by lattice-resolved (scanning) transmission electron microscopy (STEM) and electron diffraction experiments, corroborated by atomic force microscopy (AFM). Our experiments show the relative number density of graphene-nanoflakes can be tuned by optimizing the synthesis conditions. Since in the applied process the formation of the structures take place in gas-suspension, the as-synthesized CNT-G-H films can be deposited on any surface in ambient temperature with an arbitrary thickness. Moreover, this process of CNT-G-H synthesis with strong universality has also been realized in multiple systems of ethylene-based FCCVD with various catalysts and set-ups.

Carbon nanotubes (CNTs) and graphene as the typical nanocarbon materials have great potential in a wide variety of applications. CNTs, one-dimensional (1D) tubular nanomaterial have structure dependent electronic properties.[1] Depending on their cutting direction and size of the graphene sheet from which they are constructed, they can be either semiconducting or metallic.[2,3] In contrast, their parent material graphene – a two-dimensional (2D) layer of hexagonally bound carbon atoms – is a semimetal with a zero band-gap.[4] Indeed, their influence by untapping the application-potential of nanotechnology has been tremendous.[1,3,5] Recently the balance has shifted from specific molecules towards their van der Waals (vdW)[6] and covalent heterostructures.[7]

---

[*] Corresponding author.
E-mail addresses: esko.kauppinen@aalto.fi (Esko I. Kauppinen), qiang.zhang@aalto.fi (Qiang Zhang)

Interestingly, a combination of 1D-CNT and 2D-graphene heterostructure (CNT-G-H), due to synergistic effects, has shown the unique properties and even superior functional properties compared to their individual counterparts[8], in some specific applications, for example energy storage [9–11] and photonics [12,13]. Whereas, the advances in the synthesis and scalable manufacturing of these nanomaterials remain critical for tremendous influence in the application-potential of nanotechnology.[1,3,5] Therefore, recently extensive research efforts have been devoted to the synthesis of high-quality CNT, graphene as well as CNT-G-H.[10,12]

In this regard, up to the date a number of techniques, including CNT-graphene layer by layer deposition,[14] chemical vapor deposition (CVD) growth of CNTs on graphene layers [10,15,16] and liquid-based chemical synthesis methods have been utilized.[9,17] All of the above mentioned growth processes for CNT-G-H are substrate-supported, multi-step methods.[9,10,14–17] Starting from i) catalyst preparation for graphene and CNTs followed by ii) a sequence of steps for the synthesis of each graphene, CNTs and their CNT-G-H [10,15,16] and finally, iii) multi-step processing for transferring as-grown CNT-G-H from the substrate for further characterizations and applications. One of the major drawbacks of these multi-step synthesis processes is that they are highly time and resource consuming and are batch process. Moreover, the techniques involving separated growth of graphene and CNTs and then their solely physical mixing cannot provide covalent C-C bonding in between graphene and CNTs.[8,9,14,17] On the other hand, CVD growth of CNTs on graphene can provide some possibility of covalent C-C bonding but the stability of catalyst nanoparticles situated on graphene for CNT growth is highly challenging.[10,15,16] Another drawback of the substrate-supported synthesis processes is the interactions between as-synthesized CNT-G-H and substrate, which are very complicated and may have some effects on the growth chemistry in the presence of high temperature. More importantly, these interactions may introduce some defects in morphology and structure of pristine CNT-G-H while transferring sample from the substrate resulting in a change in properties of the final product.

Herein, we report a novel, scalable, gas-phase method for the substrate-free, continuous, single-step, and in-situ growth of CNT-G-H using a floating-catalyst CVD (FC-CVD) technique. It's worth noting that our technique provides a direct route for dry-deposition of as-produced CNT-G-H on substrates kept at ambient temperature for different applications or we can collect directly CNT-G-H films of desired thickness on low-adhesion membrane filter, which can be transferred

on the targeted substrate through well-established, room temperature, dry-press transfer technique.[2]

For CNT-G-H synthesis, the catalyst particles were formed by using spark discharge generation technique.[18] Briefly Fe (purity 99.8%) electrodes were used as the catalyst precursors, repeatedly evaporated by low-energy spark discharges created by applying 2-3 kV high-voltage across the electrodes gap in the presence of nitrogen ($N_2$ 99.995%) as a carrier gas. Evaporated material after going through nucleation-condensation phases formed nanoparticles.[18] The catalyst particles form spark discharge generator (SDG) were carried out by the $N_2$ (440-370 ccm) and were introduced into the vertical FC-CVD reactor as shown in Fig.1.

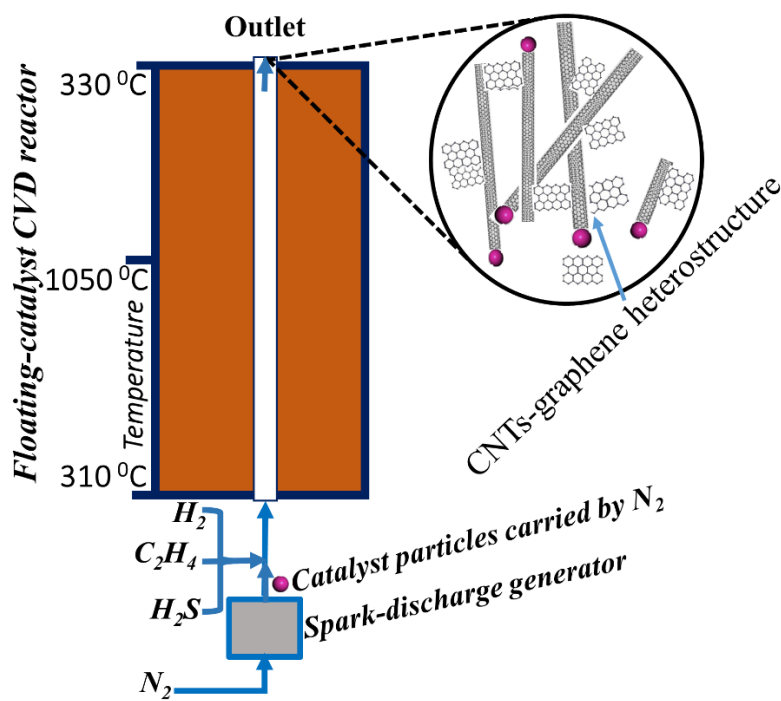

*Fig. 1. Schematic for single-step, gas-phase synthesis of CNTs graphene heterostructure (CNT-G-H) using spark discharge produced catalyst particles and vertical floating-catalyst chemical vapor deposition reactor.*

CNT-G-H was synthesized in the FC-CVD reactor using spark-produced pre-made Fe catalyst nanoparticles with 0.1sccm (200 ppm) of ethylene ($C_2H_4$) (99.999%, AGA) as a carbon source. The relative amounts of graphene and CNTs in the sample were controlled by tuning $H_2$ (99.999%, AGA) in the range 50-120 sccm at a set furnace temperature of 1050 °C. 0.01% diluted $H_2S$ (99.999%, AGA) mixed with $N_2$ having flow rate 10 sccm (corresponding to 2 ppm $H_2S$ concentration) was employed as a growth promoter. However, it's worth mentioning that the synthesis of CNT-G-H is even possible without introducing $H_2S$ but our experimental results showed that $H_2S$ has some effect on the wall conditions of the FC-CVD reactor and is helpful for maintaining long-term stability of the synthesis reactor. The residence time in the typical growth region (800-1050 °C) of the FC-CVD reactor is approximately 10 seconds and was kept constant at various $H_2$ conditions by tuning $N_2$ flow from SDG and keeping total flow 500 sccm constant inside the FC-CVD reactor.[19] At the outlet of the FC-CVD reactor as-grown CNT-G-H in the form of network were directly deposited on transmission electron microscope (TEM) copper gird for further investigations.

Scanning electron microscopy (SEM) of as-produced CNT-G-H revealed that we have graphene-nanoflakes attached with the CNTs as shown in Fig.2a. More detailed investigations of the CNT-G-H structure, were carried out by high-resolution transmission electron microscope (HR-TEM). A typical HR-TEM image of CNT-G-H is shown in Fig.2b. HR-TEM image revealed that as-produced graphene-nanoflakes in CNT-G-H are edge-enriched and have relatively small size and surface area (average size ≈ 3100 $nm^2$) which might offer rich chemistry for functionalization of dopant atoms/molecules. Moreover, our process provides in-situ, single-step growth of CNT-G-H, therefore, as-produced CNT-G-H has higher possibility of covalent C-C bonding in between graphene and CNTs.

The more direct evidence of CNT-G-H formation was obtained by atomic-resolution scanning transmission electron microscope (STEM). Fig.2c, shows a typical low magnification STEM image of as-synthesized CNT-G-H. In this image it can be seen that graphene-nanoflakes are attached with CNTs and CNTs are also providing a support to hold them. It is interesting to note that in Fig.2c, graphene-nanoflakes have different sizes and are stacked together layer by layer to form larger graphene-nanoflake. A higher magnification (atomic-resolution) STEM image of a graphene-nanoflake is provided in Fig.2d, where we have more visible lattice of graphene and CNTs. We observed that in some cases (see Fig. 2b, c and d) graphene-nanoflakes grow in the space between two CNT bundles and they are tightly bounded by CNT networks. It also indicates that we might have higher possibility of C-C bonding in between graphene and CNTs in the CNT-G-H. For further characterization, the sample of the CNT-G-H was directly collected onto mica substrate by thermophoresis technique[20] and was examined by atomic force microscope (AFM). Fig.2e and f, are typical AFM images of CNT-G-H sample and Fig.2e displays some isolated graphene-nanoflakes which are not attached with any CNT. It is obvious that the thickness and size of the nanoflakes are not uniform. In Fig.2f we can also see that some of the graphene flakes are wrapped around CNTs and some of them are making bridge between adjacent CNTs. AFM results revealed that graphene-nanoflakes not always necessarily supported by the CNTs but they can also grow separately or just detach during collection process.

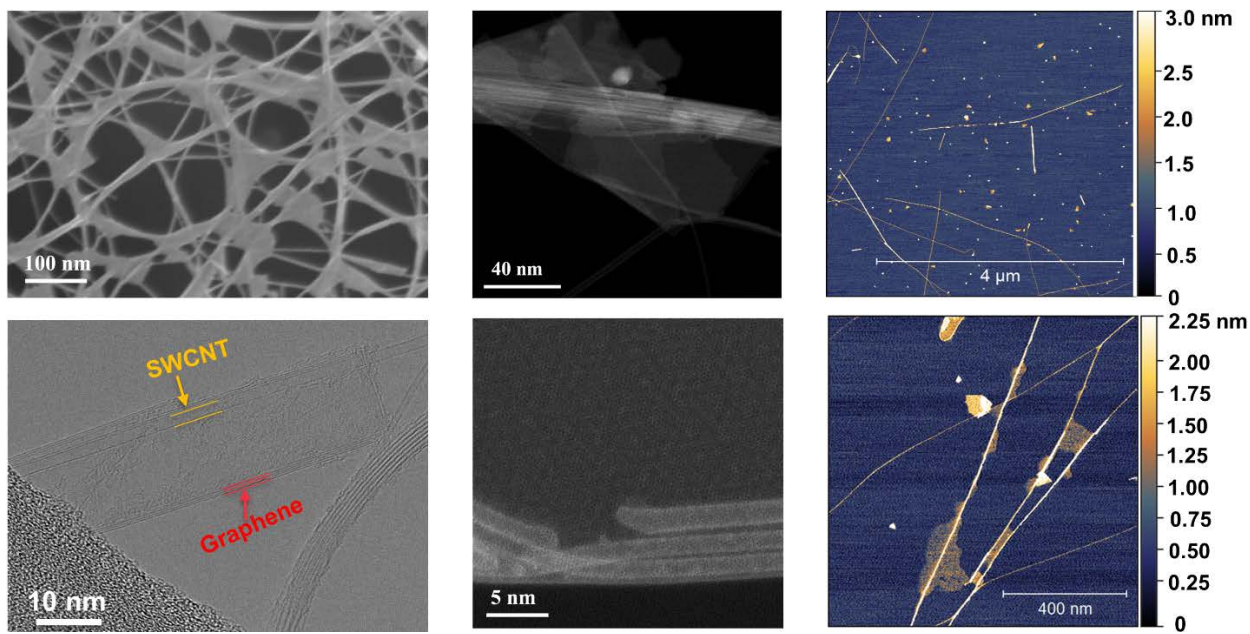

*Fig. 2. Typical images of as-synthesized carbon nanotubes and graphene nanoflakes as observed from a) scanning electron, b) high-resolution transmission electron, c) and d) scanning transmission electron, and e) and f) atomic force microscope.*

The crystallographic structure of CNT-G-H was investigated by selected area electron diffraction technique. Some of the representative CNT-G-H in Fig.3a, c and e along with their electron diffraction (ED) patterns are provided in Fig.3b, d and f, respectively. ED patterns from the CNT-G-H concurrently display features of both SWCNTs and graphene. In particular, the manifold diffraction spots arising from graphene flakes indicates an existence of two or more layers twisted into certain angles. The reason might be either our experimental conditions do not favor the growth of single-layer graphene or while moving in gas-phase from the FC-CVD reactor to the collection point, a single-layer is wrinkled to form multi-layer graphene-nanoflakes.

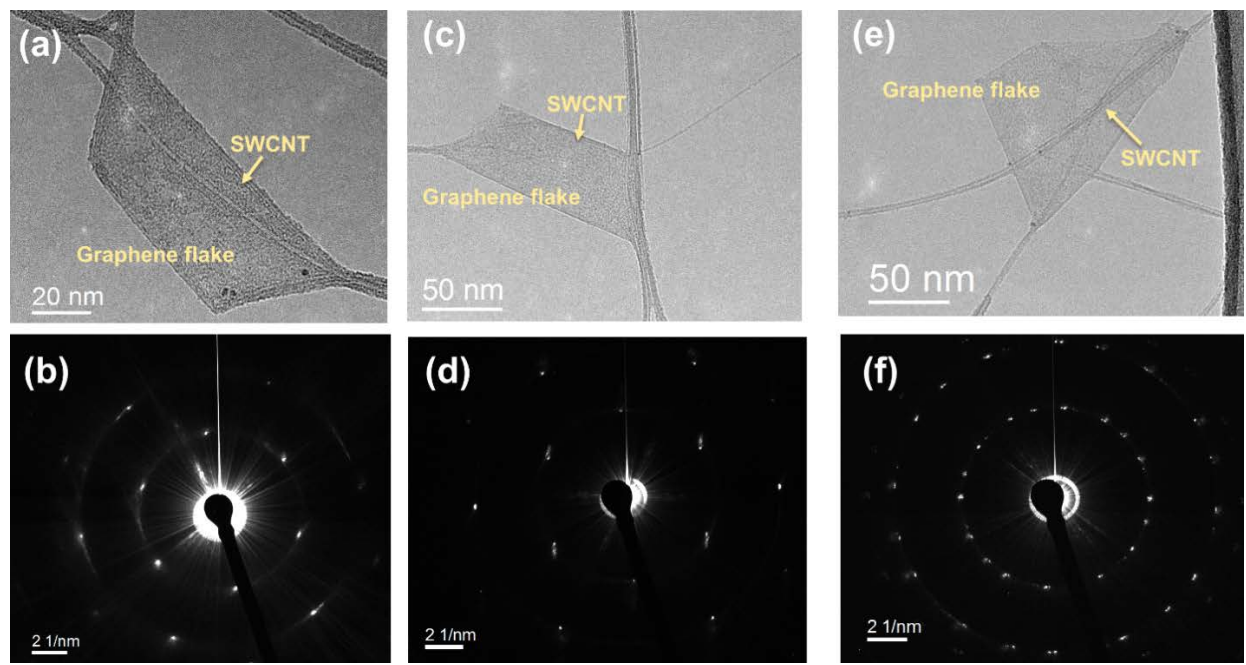

*Fig. 3. a), c) and e) High-resolution transmission electron microscope images of CNT-G-H having very sharp edges of graphene flakes and supported/attached with carbon nanotubes. b), d) and f) their electron diffraction patterns indicating features of both SWCNTs and graphene (double or multi-layers).*

The CNT-G-H synthesis is unexpected here, compared with SWCNT synthesis via FCCVD. As we known, the substrate is essential in existing methods of graphene synthesis, e.g. CVD synthesis on Cu foils[21] or epitaxial growth on Ge(110) wafer.[22] We propose the mechanism that the active carbon species directly add to the open edge sites of graphene for gas-phase CNT-G-H synthesis. The active carbon species are from the pyrolysis of hydrocarbon e.g. $C_2H_4$, which can be affected by $H_2$ concentration. Interestingly, we do find that by tuning $H_2$ flow in the FC-CVD reactor, the relative abundance of graphene-nanoflakes can be controlled. As shown in Fig.4 a-c, increasing $H_2$ from 50 sccm to 120 sccm significantly change the amount of graphene-nanoflakes in the samples. The highest amount of graphene is from the 80 sccm $H_2$, whereas the graphene will disappear when $H_2$ flow over 120 sccm.

Moreover, UV-Vis-NIR and Raman spectroscopy techniques were used for optical characterizations of CNT-G-H. Optical absorption spectra (OAS) in the wavelength range of 200-2500 nm and transmittance (%) (at 550 nm wavelength) were measured from UV-Vis-NIR spectrometer (Agilent Carry 5000; Agilent Technologies, Inc.). Raman spectrometer (Horiba

Labram-HR 800; Horiba Jobin-Yvon) was employed to acquire Raman spectra of graphene-SWCNTs by the excitation wavelength of 633 nm.

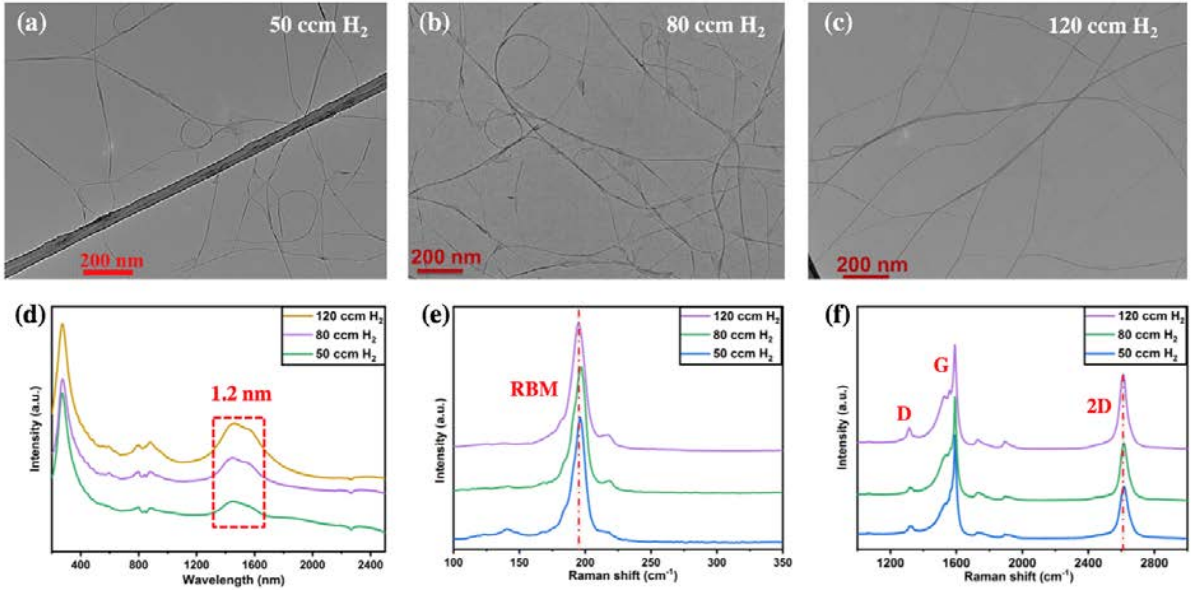

*Fig. 4. TEM images and optical characterizations of as-synthesized CNT-G-H network with variation of hydrogen amount. (a, b, c) TEM images demonstrate the dramatical change of relative number concentration of graphene nanoflakes with $H_2$ amount increase from 50 to 120 ccm. But no obvious difference in OAS (d) and Raman spectroscopy (e, f) is shown.*

Although the relative abundance of graphene-nanoflakes is very sensitive to the $H_2$ amount, but no obvious difference was found in optical characterizations as shown in Fig. 4d-f. Based on the Kataura plot and optical absorption spectra (OAS), the mean diameter of SWCNTs was calculated as 1.2 nm for $H_2$ flow of 50 to 120 sccm, which is consistent with results of Raman spectra. Radial breathing mode at 195 cm$^{-1}$ with 633 nm laser shows no obvious shift with $H_2$ amount variation. As we known, the G band originates from in-plane vibration of sp$^2$ carbon atoms and the 2D band originates from a two-phonon double resonance Raman process.[23] Specially, the 2D band is closely related to the band structure of graphene layers. As a result, the presence of a sharp and symmetric 2D band is widely used to identify single layer of graphene. When the graphene thickness increases, the G-band intensity increases almost linearly and the 2D band becomes broader and blue shifted. Moreover, the differences in the 2D band between two and few layers of graphene are not unambiguous in the Raman spectra.[24] For all samples with 50, 80 and 120 sccm $H_2$, the symmetric peaks of 2D mode are at ~ 2610 cm$^{-1}$ and intensity ratios of 2D mode to G mode are ~ 0.6 to 0.8. The reasons might be most graphene flakes over 2 layers and low graphene content. The highest graphene content at 80 sccm $H_2$ is around 5% in graphene-SWCNT samples.

Moreover, the synthesis of CNT-G-H also has been realized in other ethylene-based FCCVD system with various catalysts including Co, Co-Ni catalysts from spark discharge[18], and Fe catalysts from Ferrocene decomposition[25] by precise control of growth parameters. The typical graphene-SWCNTs samples from ferrocene-$C_2H_4$ system are shown in Fig. 5. AFM image shows the samples contain both CNT-G-H and isolated graphene-nanoflakes. High-resolution STEM image of the hexagonal lattice clearly indicates the as-synthesized nanoflake is graphene. These results suggest this process has strong universality and also support the proposed mechanism that the active carbon species from hydrocarbon pyrolysis directly add to the open edge sites for gas-phase graphene synthesis.

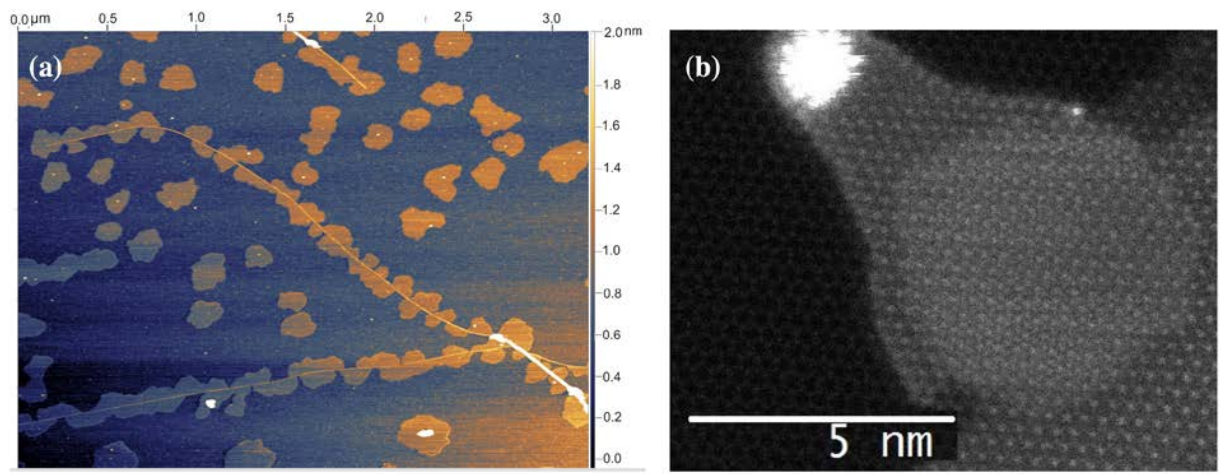

*Fig. 5. Typical AFM and STEM images of graphene-SWCNTs from the ferrocene-ethylene FCCVD system. (a) AFM image showing the samples containing both isolated graphene-nanoflakes and CNT-G-H. (b) High-resolution STEM image of the hexagonal lattice clearly indicates graphene has been synthesized.*

In summary, we have developed a single-step, in-situ FC-CVD growth process for the free-standing CNT-G-H, by using ethylene as a carbon source. Both the electron diffraction patterns and atomic-resolution STEM images of graphene-nanoflakes and CNTs confirmed the formation of CNT-G-H. HR-TEM images revealed that as-synthesized graphene-nanoflakes are edge-enriched and might offer more sites for functionalization of dopant atoms/molecules. The relative number density of graphene-nanoflakes can be optimized in the CNT-G-H by tuning $H_2$ amount in the FC-CVD synthesis reactor. Moreover, this process of CNT-G-H synthesis with strong universality has been realized in multiple systems of ethylene-based FCCVD with various catalysts and set-ups. This method is purely in gas-phase and has potential to be scaled-up for the production of high-quality CNT-G-H at industrial scale.